\documentclass[prl,twocolumn,showpacs,preprintnumbers,amsmath,amssymb]{revtex4}
\usepackage{amsmath,epsfig,graphicx,amssymb}
\begin{document}

%\preprint{APS/123-QED}
%\tightenlines
% \draft
\title{Two-Photon Excitation of Low-Lying Electronic Quadrupole States
  in Atomic Clusters}

\author{V.O. Nesterenko$^1$, P.-G. Reinhard$^2$, T. Halfmann$^3$ 
and L.I. Pavlov$^4$}
\date{\today}
\affiliation{$^{1}$Laboratory of Theoretical Physics,
Joint Institute for Nuclear Research, Dubna, Moscow region, 141980,
Russia}
\email{nester@theor.jinr.ru}
\affiliation{$^{2}$ Institute of Theoretical Physics II,
University of Erlangen-Nurnberg, D-91058, Erlangen, Germany}
\affiliation{$^{3}$ Department of Physics, Technical University
Kaiserslautern, D-67653, Germany}
\affiliation{$^{4}$ Institute of Electronics, Bulgarian Academy
of Sciences, Sophia, Bulgaria}

%\date{today}

\begin{abstract}
A simple scheme of population and detection of low-lying
electronic quadrupole modes in free small deformed metal clusters
is proposed. The scheme is analyzed in terms of the TDLDA
(time-dependent local density approximation) calculations. As test
case, the deformed cluster $Na^+_{11}$ is considered. Long-living
quadrupole oscillations are generated via resonant two-photon
(two-dipole) excitation and then detected through the appearance
of satellites in the photoelectron spectra generated by a probe
pulse. Femtosecond pump and probe pulses with intensities $I =
2\cdot 10^{10} - 2\cdot 10^{11} W/cm^2$ and pulse duration $T =
200 - 500$ fs are found to be optimal. The modes of interest are
dominated by a single electron-hole pair and so their energies,
being combined with the photoelectron data for hole states, allow
to gather new information about mean-field spectra of valence
electrons in the HOMO-LUMO region.  Besides, the scheme allows to
estimate the lifetime of electron-hole pairs and hence the
relaxation time of electronic energy into ionic heat.
\end{abstract}

\pacs{36.40.Cg; 42.62.Fi; 42.50.Hz}

\maketitle

%\arraycolsep1.5pt
%\narrowtext\preprint{HEP/123-qed}
%\narrowtext

In recent years, the analysis of cluster structure and dynamics has made
remarkable progress in going beyond mere optical absorption spectroscopy.  More
and more observables are now being accessed with methods which had been applied
successfully in atomic and molecular physics, as e.g. measuring angular
distributions, photoelectron spectra (PES), or pump and probe scenarios, for
an overview see \cite{Reinbook}. For example, PES have attracted much interest,
see e.g. \cite{Wrigge,Moseler,Alum,c60_cam}. They allow to determine the
energies $\epsilon_{h}$ of the occupied (hole) electron levels in the mean
field of the cluster. Altogether, cluster studies are shifting from the
investigation of global properties (like, e.g. the dipole plasmon) to exploring
more detailed features, as e.g. the single-electron states in the cluster mean
field. These novel observables are important not only for clusters but for any
nano-system since they deepen the understanding and promote the
development of experimental techniques.

In spite of impressive achievements, atomic clusters bear still a
variety of unexplored features. For example, clusters have a rich
excitation spectrum of higher multipolarities $\lambda> 1$, beyond the
well studied dipole modes.  As is discussed below, low-lying
non-collective excitations of this family give access to the
single-electron spectra and allow to estimate some principle time
scales relevant for clusters and other nano-systems. However, the
standard one-photon absorption experiments are blind to non-dipole
states. Other means of analysis as, e.g., inelastic electron
scattering did not yet come very far due to the small energies
involved. It is the aim of present contribution to discuss a possible
pathway to the particular kind of non-dipole spectra, quadrupole
($\lambda =2$) non-collective modes. The novel pump/probe technique
with using laser two-photon process will be proposed for population
and detection of these modes.

It is a challenging task as such to get access to quadrupole
modes. Moreover, the low-lying quadrupoles carry useful
information about the unoccupied electron states near the Fermi
surface.  The modes are predominantly of electron-hole ($eh$) type
\cite{Ne_PRA_2004}, whose spectra are the mere energy differences
$\epsilon_{eh}= \epsilon_{e}-\epsilon_{h}$. Knowing the hole state
$\epsilon_{h}$ from other sources, allows then to conclude on the
involved particle energies $\epsilon_{e}$.  A further interesting
observable is the lifetime $\tau_{eh}$ of an $eh$ pair since it
provides the time scale of the transfer of electronic energy into
internal energy ( mainly ionic motion and to some extend higher
correlated electron states). The value $\tau_{eh}$ is
important for many processes, e.g., for defining different regimes
of multi-photon ionization (MPI) in atomic clusters
\cite{c60_cam}. Being determined by
electron-ion correlations, $\tau_{eh}$ should be much longer than
the lifetime of the dipole plasmon. However, very little is known
experimentally. As is shown below, the scheme we propose is also
appropriate for measurement of the lifetime $\tau_{eh}$.

\begin{figure}
\includegraphics[height=7.8cm,width=7.0cm]{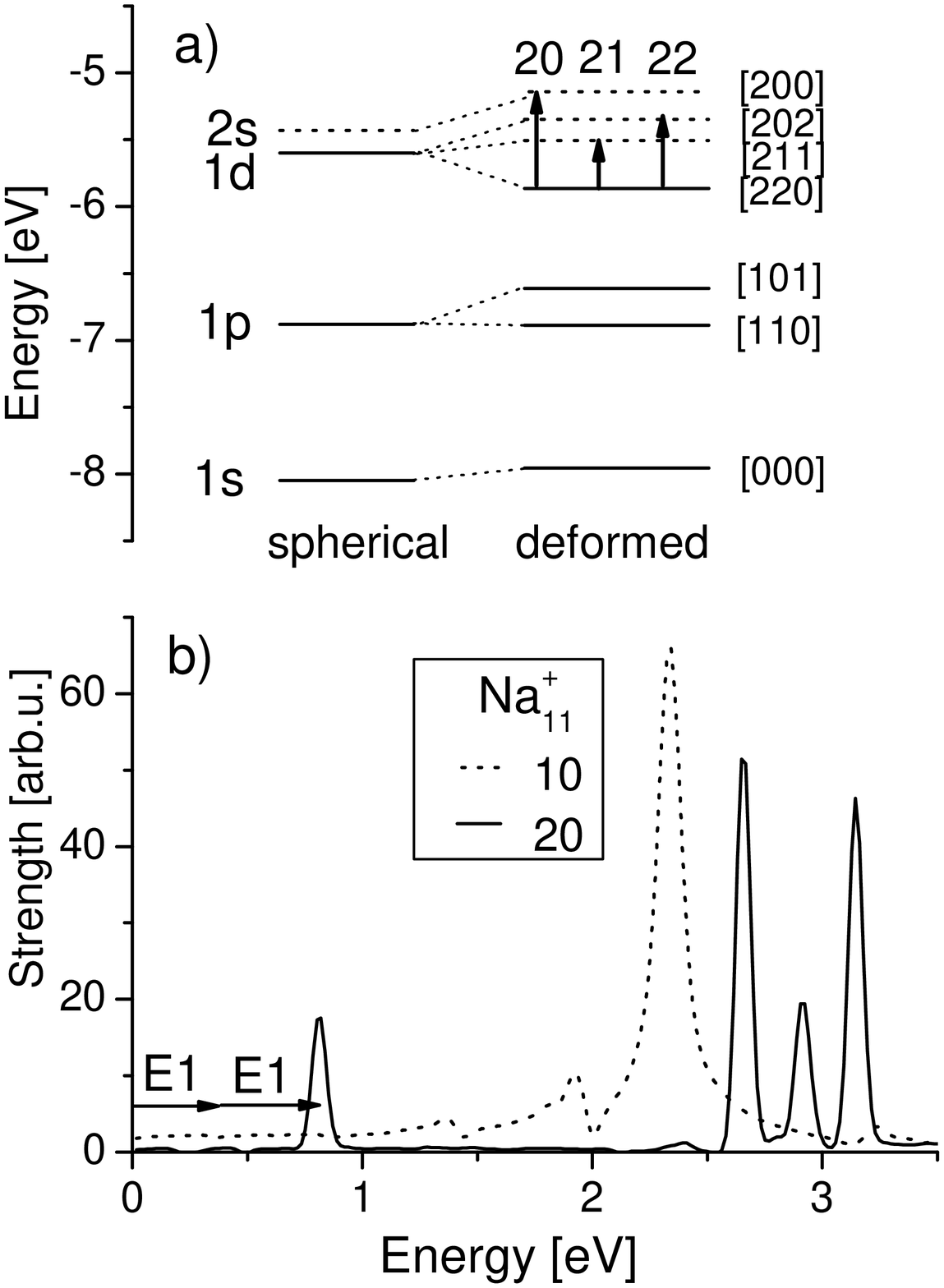}
\\
\caption{\label{fig:spectr_fig1}
a) The lowest single-electron states for Na$^+_{11}$ in
the spherical limit (left) and at equilibrium deformation (right).
The spherical levels are marked by oscillator
quantum numbers $nl$ and the deformed levels by the Nilsson-Clemenger
notation $[Nn_z\Lambda]$ \protect\cite{Clem}. Occupied and unoccupied levels
are drawn by solid or dotted lines, respectively. Arrows depict infrared
$eh$ excitations with multipolarity $\lambda\mu =$20, 21 and 22.\\
b) Dipole ($\lambda\mu =$10) and quadrupole ($\lambda\mu =$20)
strengths distributions in Na$^+_{11}$. The large peaks above 2 eV represent
the dipole and quadrupole plasmons.
The horizontal arrows depict the two-photon (two-dipole) resonant excitation
of the target quadrupole state with $\lambda\mu =$ 20.
}
\end{figure}

Low-lying (infrared=IR) electronic quadrupole states
in free small deformed clusters seem to be most appropriate for our aims
\cite{Ne_PRA_2004}. Beams of size selected small clusters are readily
available. In free small clusters, the IR electronic spectra are dilute, which
simplifies their experimental discrimination. Thus we concentrate here on these
species. Deformed clusters are needed, because most of the IR quadrupole
states arise due to the cluster deformation and are absent in spherical
clusters (see the upper plot in Fig. \ref{fig:spectr_fig1}). This then provides
a further indicator of the cluster shape. And, as outlined above, low-lying
quadrupole states in small deformed clusters give access to $eh$ energies
since  they can be, to a good approximation, considered as pure $eh$ states
\cite{Ne_PRA_2004}.

Quadrupole states cannot be populated in mere photo-absorption
where dipole modes dominate. One has to use two-photon
(two-dipole) processes. An option is one of the different versions
of the stimulated Raman process (e.g., off-resonant Raman it{or}
stimulated Raman adiabatic passage (STIRAP) \cite{Berg}) which
steps up via an intermediate dipole state. These methods use three
laser pulses: pump, Stokes and probe.
They may be applicable to atomic clusters when choosing appropriate
laser parameters \cite{recent}.

In the present paper, we will consider an alternative, probably
simpler, scheme using direct two-photon excitation in resonance with
the final quadrupole state
(see Fig.\ref{fig:spectr_fig1}b)).  To detect the successful
population of the quadrupole states, the cluster is irradiated by
a probe pulse (with appropriate delay) leading to MPI. The
corresponding PES are recorded. The quadrupole oscillation couples
with the single-electron PES structures and thus manifests itself
via satellites in the PES. In this scheme we need only two lasers,
pump and probe. As is shown below, intense fs lasers are most
appropriate.

As a test case for our computational simulation, we consider Na$_{11}^+$ in
jellium approximation. This cluster is strongly prolate and displays
well separated low-lying quadrupole states. The calculations are
performed within the time-dependent local density approximation (TDLDA) with
the exchange-correlation functional \cite{ex_corr}.  The ions are treated in
the soft jellium approximation \cite{Reinbook}. This approximation,
although a bit daring to small metal clusters, is well suited for the
exploration of the two-photon method in $Na^+_{11}$ which is axially symmetric
in jellium. Besides, this allows to perform the
calculations on an axial grid in coordinate space. Absorbing boundary
conditions are employed for the description of photoionization.  
%The numerical handling is done by standard methods
%(gradient iteration for the ground state, time splitting of propagation, SIC
%correction), for details see \cite{Cal,Reinbook}. 
The excitation spectra in the
linear response regime are computed using TDLDA by standard techniques of
spectral analysis \cite{spectran,Cal}. The laser induced dynamics is simulated
by adding to the TDLDA the laser pulses (propagating in z-direction)
as classical external dipole fields of
the form $W(t)=E_0\,z\,\sin^2(t/T)\cos (\omega t)$ where $E_0$ is the
field strength ($\propto$ square root of the intensity $I$), $\omega$ the
frequency, and $T$ the pulse duration \cite{Cal,Reinbook}.
%
%%%%%%%%%%%%
% Figure 1
%%%%%%%%%%%%

Fig. \ref{fig:spectr_fig1} demonstrates the spectra of $Na_{11}^+$.
Panel a) shows the single-electron states.  The ground state is
deformed. Comparison with the spherical spectrum shows that the axial
deformation splits the single-electron spectra and hence the energies
of $eh$ pairs.  Following the scheme, only three low-lying quadrupole
$eh$ excitations, $\{ [220],[200]\}_{20}$, $\{ [220],[211]\}_{21}$,
and $\{ [220],[202]\}_{22}$, exist in $Na_{11}^+$ and the last two of
them are fully driven by the cluster deformation. The associated
excitation energies are small because they are mainly determined
by the deformation splitting.  Panel b) shows the relevant part of
the excitation spectrum in terms of the dipole
($\lambda\mu=10$) and quadrupole ($\lambda\mu=20$) strengths. The
modes with $\mu > 0$ (and thus the excitations $\lambda\mu=21$ and
22) are not presented here.
To illustrate our method it is quite enough to consider only
one low-energy quadrupole mode, namely that of  $\lambda\mu=$20.
This mode is seen as a peak at the energy $e_{20}=$0.8 eV.

Figure 1 shows that the only habitants of the IR region below
1.3 eV are three low-energy quadrupole excitations indicated by arrows
in the panel a). Both dipole and quadrupole plasmons lie well above.
The lowest dipole $eh$ excitation also resides above, at 1.4 eV. So, the
IR part of the spectrum we are interested in is indeed very
dilute. This favors the experimental discrimination of the quadrupole
levels. Moreover, the well separated levels are not subject to
collective mixing and thus preserve their $eh$ nature.

The lifetime of the low-lying quadrupole states probably does not
exceed  several ps. The exciting laser pulses should be
shorter, say hundreds fs. They have to be intense
enough to excite the quadrupole mode in the two-photon process and
to yield a measurable PES. On the other hand, the pulses should
not be too intense in order to keep the resolution in the PES.
As is shown below, the optimal intensity is $I=2\cdot 10^{10}-
2\cdot 10^{11} W/cm^2$. The simplest case of one-photon (n=1)
ionization cannot be applied here, since the low-order
ionization would demand lasers operating in the ultraviolet (UV).
But UV lasers with sufficient intensities are available only in the ns
regime which is useless for our purposes. So, we are enforced to deal
with fs lasers and multi-photon ionization (MPI). To minimize the
MPI order $n$, it is desirable to work at the minimal wavelength
available from fs laser systems. In the present study, we use the laser
radiation at $\hbar\omega_{probe}=3.2$ eV, which corresponds to the second
harmonic frequency of a titanium sapphire laser. The MPI order is then $n=2$.
The tunable IR pump pulse may be provided by fs optical-parametric generators.
%
%%%%%%%%%%%%
% Figure 2
%%%%%%%%%%%%
\begin{figure}
\includegraphics[height=8cm,width=7cm,angle=-90]{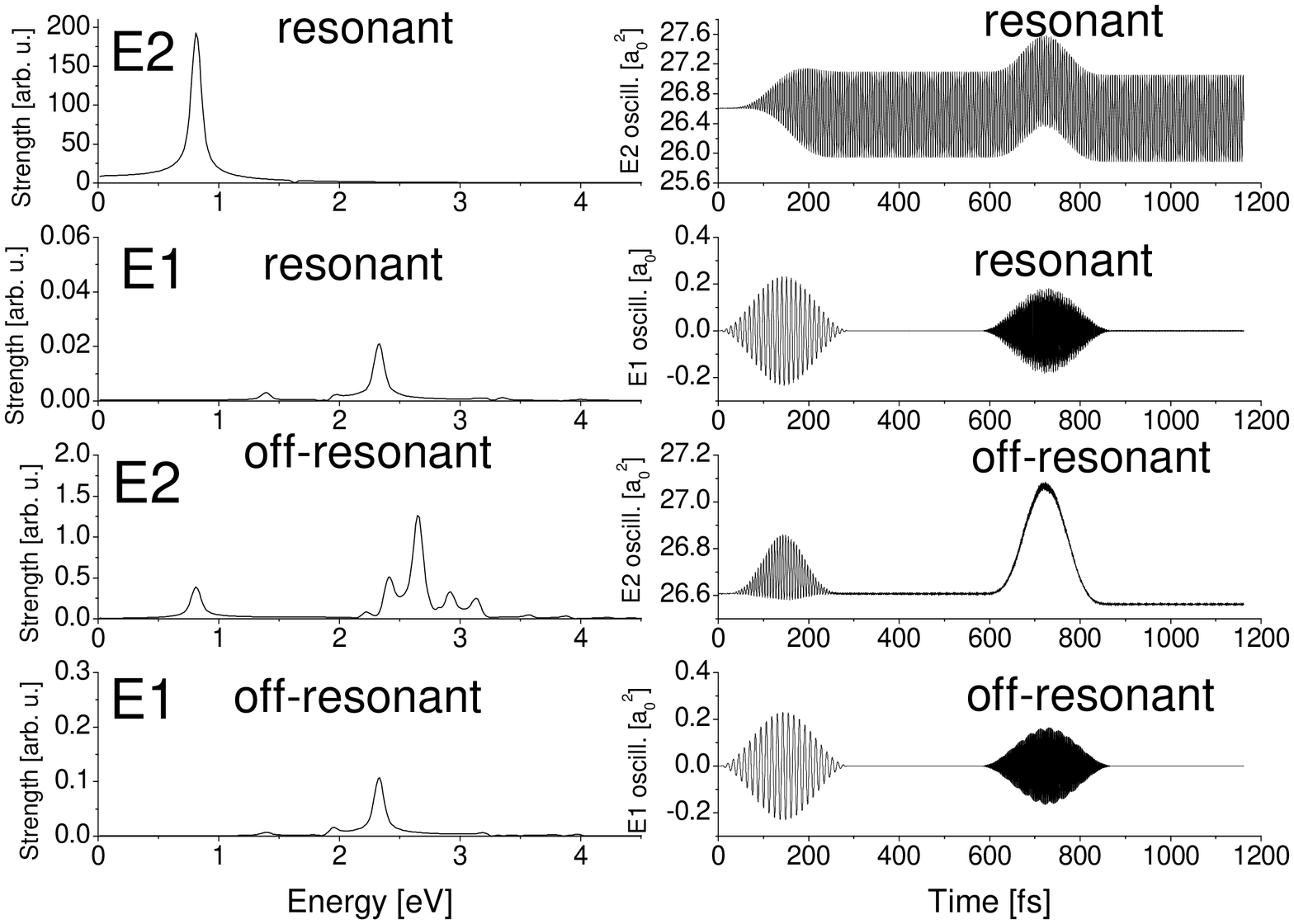}
\caption{\label{fig:e1_e2_fig2}
Resonant and off-resonant
excitations in Na$^+_{11}$. The left panels exhibit quadrupole and dipole
strengths as function of the excitation energy.  The right panels depict
the electronic quadrupole and dipole moments (in
atomic units) as a function of time. The pump (first) and probe
(second) pulses are clearly seen. The
calculations are performed for laser intensities
$I_{pump}=I_{probe}=10^{11} W/cm^2$ and pulse durations
$T_{pump}=T_{probe}= 300 fs$.  }
\end{figure}

Fig. 2 exhibits the first step of our scheme, the population of
the $\lambda\mu = 20$ quadrupole state.  The right panels shows
time evolution of the dipole and quadrupole oscillations caused by
the pump pulse. The resonant ($2\hbar\omega_{pump}=2\cdot 0.40$ eV
$= e_{20}=0.80$ eV) and off-resonant ($2\hbar\omega_{pump}=2\cdot
0.34$ eV $\ne e_{20}=0.80$ eV) cases are considered.  It is
obvious that only the resonant case develops self-sustaining
quadrupole oscillation. Since electron-ion relaxation is not taken
into account here, this oscillation persists for several ps and
further. But the dipole oscillates only during the pump pulse at t
= 0 - 300 fs. In the off-resonant case, both quadrupole and dipole
modes respond only during the pump pulse.  The left panels exhibit
the corresponding dipole and quadrupole strengths. Note that
in the resonant case (left upper panel), the quadrupole
mode of interest at 0.80 eV dominates all other modes. This
resonant suppression of competitors is crucial for producing the
clean satellites in PES later on.
The quadrupole mode as excited by the resonant two-photon
excitation continues to oscillate for a while.
The population of this mode can thus be checked by a
sufficiently delayed probe pulse. This pulse is shown
at t = 600 - 900 fs. The pulse delay 600 fs (equal to the double
pulse duration) suffices to all other (virtual) modes to relax to
zero, except of the long-living mode of interest.
%%%%%%%%%%%%
% Figure 3
%%%%%%%%%%%%
\begin{figure}
\includegraphics[height=4.5cm,width=6.5cm]{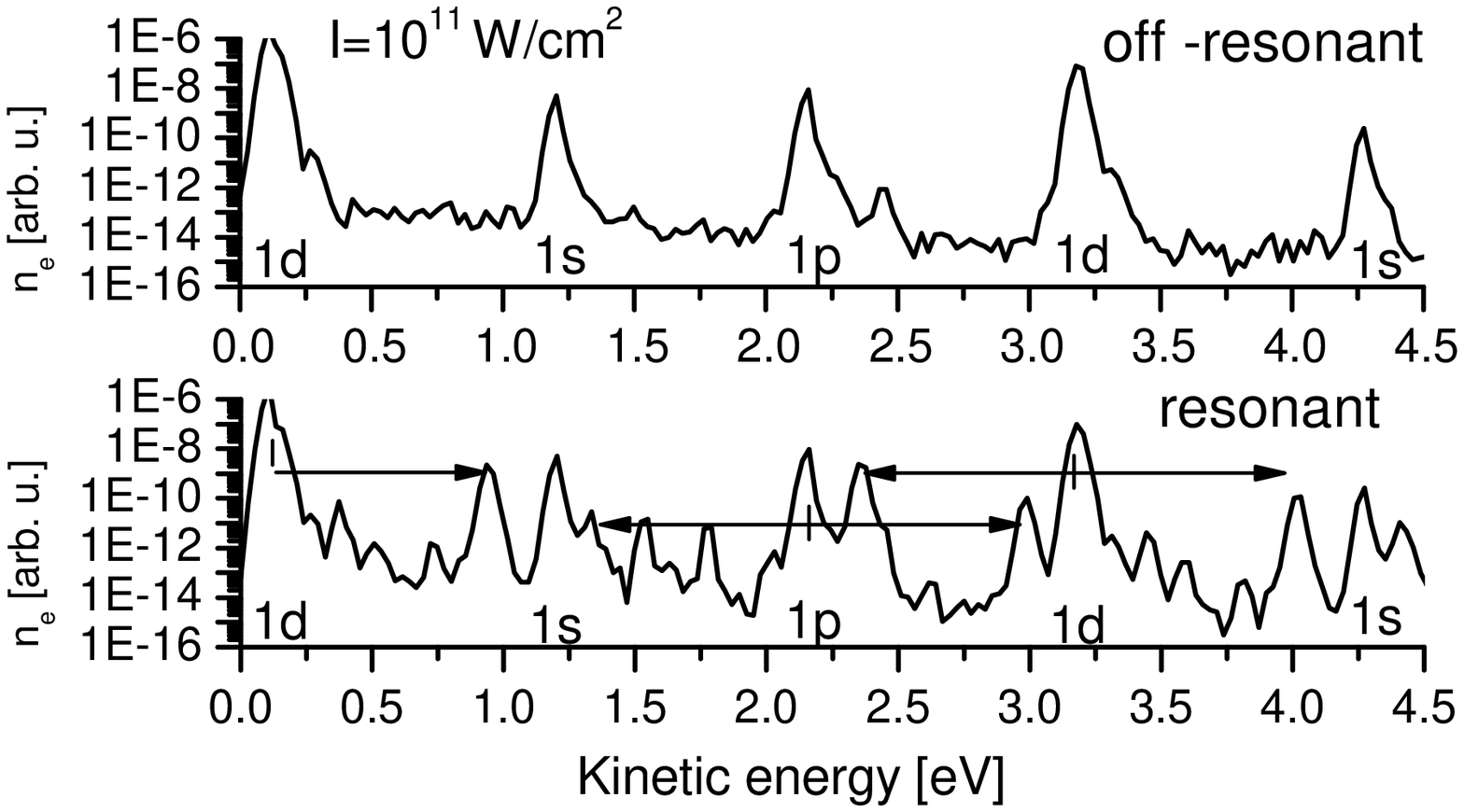}
\caption{\label{fig:ne_in_off_fig3}
The photoelectron yield in Na$^+_{11}$ at time t=1200 fs
for the off-resonant and resonant cases.
Parameters of the calculations are the same as in Fig. 2.
}
\end{figure}

The probe pulse leads to strong photoionization and the resulting PES
provides information about the cluster state before ionization. The upper
panel of Fig. 3 shows the PES for the off-resonant case, which
is typical for the cluster ground state
\cite{pohl}. The structures corresponding to single-electron hole
states $1d$, $1p$ and $1s$ are well visible.  The lower panel shows
the PES for the resonant case.  One recognizes additional satellites
in the spectra which are shifted just by the quadrupole energy $\pm
0.8$ eV with respect to the leading peaks, as indicated by the arrows. They
emerge from the coupling of the quadrupole mode with the
single-electron states. Only the strongest $1d$ and $1p$ PES
structures have significant satellites. The most remarkable satellite
is at $e_{kin}\sim$ 1 eV.
The interpretation of the quadrupole energy in terms of $eh$
differences $\epsilon_{eh}= \epsilon_{e}-\epsilon_{h}$
is to be taken with a bit of care.  The state can involve
a small coupling to collective quadrupole strength. Detailed
analysis shows that the $\lambda\mu$=20 mode in Fig. 2 contains a
small collective blue-shift of $< 0.1$ eV as compared with the energy of
the pure $\{ [220],[200]\}_{20}$ configuration. However, this shift
is within the accuracy of the typical PES measurements.

The proposed scheme allows to obtain not only the frequency of the
quadrupole mode but also its lifetime. To that end, one
should simply increase, step by step, the delay between the pump and
probe pulses. The relaxation of quadrupole oscillation will finally
lead to an extinction of the satellites from which one can read off
the lifetime. This feature cannot be tested in our present model
where $eh$ states have infinite lifetime.

In Fig. 4 the sensitivity of the process to the pulse intensity is
tested. The PES structures and their satellites are strong and can
be well discriminated for intensities $I=2\cdot 10^{10}- 2\cdot
10^{11} W/cm^2$. Smaller intensities result in too weak
photoelectron yield while higher intensities lead to considerable
broadening the PES (see also
\cite{pohl}). The calculations with the fixed intensity $I=
10^{11} W/cm^2$ (as in Figs. 2 and 3) but different pulse
durations show that values of $T = 200-500 fs$ are most optimal.
Altogether, the optimal intervals for pulse intensities and
durations are comfortably broad which is a promising feature for
future experiments.
%%%%%%%%%%%%
% Figure 4
%%%%%%%%%%%%
\begin{figure}
\includegraphics[height=9.0cm,width=7.5cm]{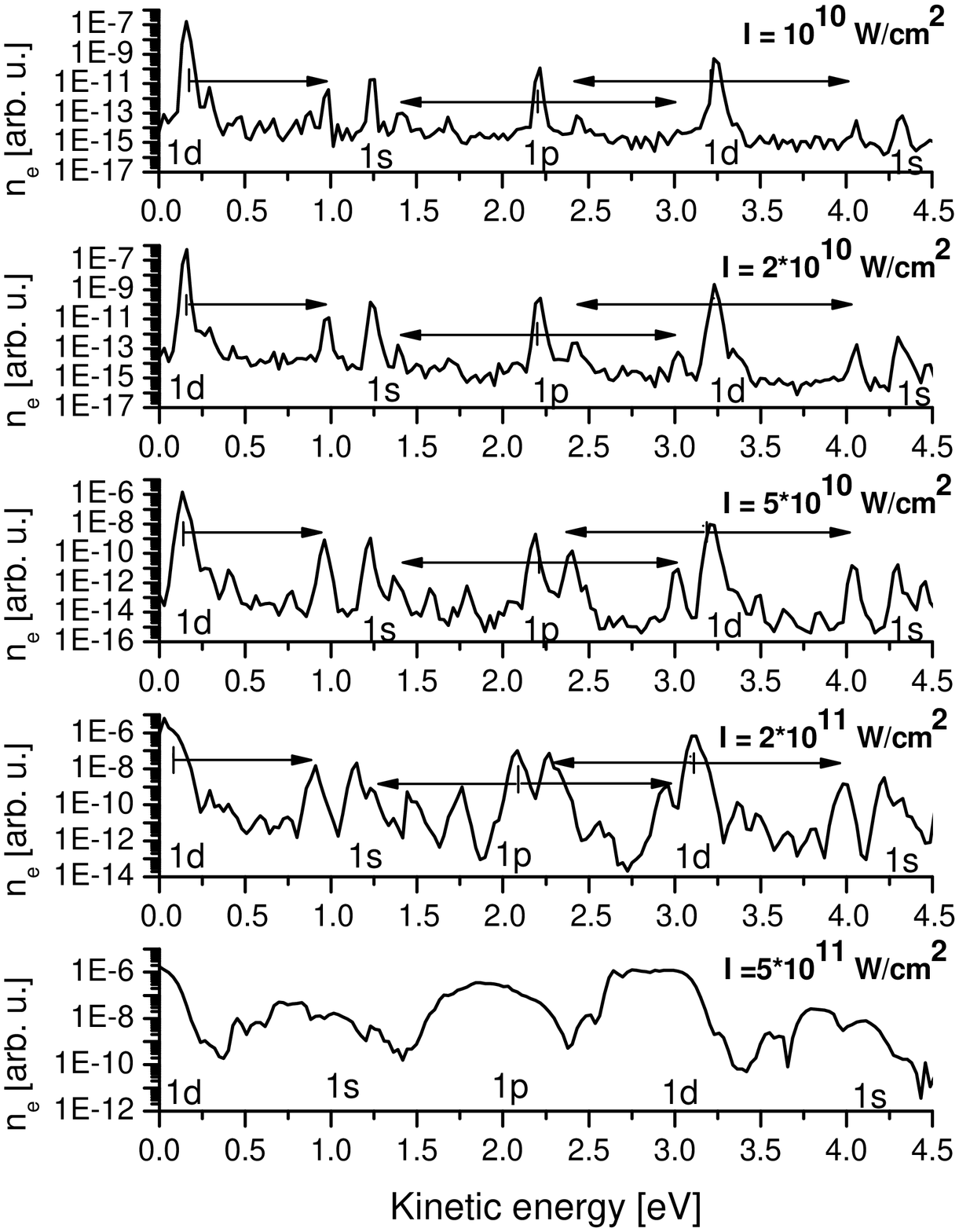}
\caption{\label{fig:vary_int_fig4}
The resonant photoelectron yield in Na$^+_{11}$
for different laser intensities as indicated.
All other parameters are the same as in Figs. 2 and 3.
}
\end{figure}

Let us finally outline the proposed experiment.
One needs a synchronized tunable infrared pump pulse and UV
probe pulse. The optimum probe parameters are found by
scanning the intensities while looking at the PES from the ground
state. Here our predictions for the optimal pulse intensities and
durations can be used as a first guide. The largest intensity
which still produces well discriminated peaks is to be chosen. A
similar intensity will be appropriate for the pump pulse (although
separate tuning may improve the results). One then scans the pump
frequency for a probe pulse delay of order of the double pulse
length and watches the appearance of the
satellites. The maximum satellite signal provides the quadrupole
energy in two ways, first as the double pump frequency and second, as a
countercheck, from the offset of the satellites.
In a last step, one can vary the pulse delay to find out the lifetime of
the quadrupole state.

The low-energy spectra
are sensitive to thermal fluctuations of the cluster configuration.
Our survey deals with a fixed ground state configuration. Thus it
is advisable to work at low temperatures (less 100 K)
to avoid broadening of the signal and other artifacts. Our
calculations show that the photoelectron yield from the satellites
should be $n_e \sim 10^{-5}$ electrons per cluster and laser
pulse. Then, assuming typical parameters of available cluster beams
\cite{Issen} and lasers with kHz repetition, one finds that $n_e$ still
should be measurable even under additional damping
effects (electron-ion correlations, possible probe-induced multi-plasmon
excitations \cite{Pohl_2}, random orientation of
clusters in the beam).

In conclusion, we propose a simple scheme for the population and
detection of non-collective (electron-hole) infrared quadrupole states
in small deformed atomic clusters. It relies on a pump-and-probe
technique where the pump pulse excites the quadrupole mode via a
two-photon process and the probe pulse explores the resulting
quadrupole oscillations through satellites in the photoelectron
spectra. Already the measurement of low-lying quadrupole states in
itself is an achievement as such. Moreover, it can deliver
information about the single-electron spectrum directly above the
Fermi energy and about the relaxation time for the electron-hole
pairs. The proposed scheme is quite general and can be applied
to any clusters provided their electronic infrared spectrum is
sufficiently dilute.

%\vspace{0.2cm} {\bf 
\begin{acknowledgments} 
V.O.N. thanks Prof. K. Bergmann for
valuable discussions and hospitality during the stay in Kaiserslautern
university. The work was partly supported  by the DFG
(project GZ:436 RUS 17/104/05)
and Heisenberg-Landau (Germany-BLTP JINR) grants.
\end{acknowledgments}


\begin{references}
\bibitem{Reinbook} %1
    P.-G. Reinhard and E. Suraud,
    {\em Introduction to Cluster Dynamics},
    (Wiley-VCH, Berlin, 2003).
\bibitem{Wrigge} %2
   G. Wrigge, M. Astruc Hoffmann and B.v. Issendorff,
   Phys. Rev. A{\bf 65}, 063201 (2002).
\bibitem{Moseler} %3
   M. Moseler, et al,
   %B. Huber, H. H${\ddot a}$kinen, U. Landman, G. Wrigge, M. Astruc Hoffmann and B.v. Issendorff,
   Phys. Rev. B{\bf 68}, 165413 (2003).
\bibitem{Alum} %4
   J. Akola, et al,
   %M. Manninin, H. Hakkinen, U. Landman, X. Li, and L.-S. Wang,
   Phys. Rev. B{\bf 60} R11297 (1999).
\bibitem{c60_cam} %5
      E.E.B. Campbell, et al,
      %K. Hoffmann, H. Rottke, and I.V. Hertel,
      J. Chem. Phys. {\bf 114}, 1716 (2001).
\bibitem{Ne_PRA_2004} %6
     V.O. Nesterenko, P.-G. Reinhard, W. Kleinig and D.S. Dolci,
     Phys. Rev. A{\bf 70}, 023205 (2004).
\bibitem{Berg} %7
     K. Bergmann, H. Theuer, and B.W. Shore,
     Rev. Mod. Phys. {\bf 70}, 1003 (1998);
     N.V. Vitanov, et al,
     %M. Fleischhauer, B.W. Shore, and K. Bergmann,
     Adv. Atom. Mol. Opt. Phys., {\bf 46}, 55 (2001).
\bibitem{recent} %8
     V.O. Nesterenko, P.-G. Reinhard, Th. Halfmann,
     under preparation for publication.
\bibitem{ex_corr} %9
     J.P. Perdew and Y. Wang,
     Phys. Rev. B{\bf 45}, 13244 (1992).
\bibitem{Cal} %10
    F. Calvayrac, et al,
    %P.-G. Reinhard, E. Suraud, and C.A. Ullrich,
    Phys. Rep. {\bf 337}, 493 (2000).
\bibitem{spectran} %11
     F. Calvayrac, et al,
     %E. Suraud, and P.-G. Reinhard,
     Ann. Phys. {\bf 254} (N.Y.), 125 (1997)
\bibitem{Clem} %12
     K. Clemenger, Phys. Rev. B{\bf 32}, R1359 (1985).
\bibitem{pohl} %13
     A. Pohl, P.-G. Reinhard, and E. Suraud,
      Phys. Rev. Lett. {\bf 84}, 5090 (2000)
\bibitem{Issen} %14
     B. v. Issendorf, private communication.
\bibitem{Pohl_2} %15
   A. Pohl, P.-G. Reinhard, and E. Suraud,
   J. Phys. B{\bf 34}, 4969 (2001).

\end{references}
\end{document}